\documentclass[prb,aps,showpacs]{revtex4}
\usepackage{graphicx}
\usepackage{subfigure}
\usepackage{amstext,amsfonts,amsbsy,amssymb,bbm}
\usepackage{array,multirow}
\usepackage[pageshow]{supertabular}
\usepackage{color}
\usepackage{setspace}

\newcommand {\be}[1]{\begin{eqnarray} \mbox{$\label{#1}$}  }
\newcommand{\ee}{\end{eqnarray}}

\newcommand{\pref}[1]{(\ref{#1})}

\newcommand\ie {{\it i.e.}, }

\newcommand{\nn}{\nonumber\\}

\newcommand\half{\frac 1 2 }

\newcommand{\hf}{ {\frac{1}{2}}}

\newcommand{\pd}{\partial}

\newcommand{\pdn}[1]{\partial_{#1}}

\newcommand{\mean}[1]{\left \langle #1 \right \rangle}

\newcommand{\gb}{ {\beta} }

\newcommand{\gd}{ {\delta} }
\newcommand{\gD}{ {\Delta} }

\newcommand{\gw}{ {\omega} }

\newcommand{\gr}{\rho}

\newcommand{\gQ}{ {\Theta} }

\begin{document}

\title{Charge Fractionalization on Quantum Hall Edges}
\author{Mats Horsdal$^{a,b}$, Marianne Rypest{\o}l$^{c}$, Hans Hansson$^{d}$  and Jon Magne Leinaas$^{c}$}
\affiliation{${(a)}$ Institute for Theoretical Physics, University of Leipzig, D-04009 Leipzig, Germany  }
\affiliation{${(b)}$ Nordita, Roslagstullsbacken23, SE-106 91Stockholm, Sweden   }
\affiliation{${(c)}$ Department of Physics, University of Oslo, N-0316 Oslo, Norway}
\affiliation{${(d)}$ Department of Physics, Stockholm University, AlbaNova University Center, SE-106 91 Stockholm, Sweden}
 
\date{September 22, 2011}
\begin{abstract}
We discuss the propagation and fractionalization of localized charges on the edges of quantum Hall bars of variable widths, where interactions between the edges give rise to Luttinger liquid behavior with a non-trivial interaction parameter $g$. We focus in particular on the separation of an initial charge pulse into a sharply defined front charge and a broader tail. The front pulse describes an adiabatically dressed electron which carries a non-integer charge, which is $\sqrt g$ times the electron charge. We discuss how the presence of this fractional charge can, in principle, be detected through measurements of the noise in the current created by tunneling of electrons into the system. The results are illustrated by numerical simulations of a simplified model of the Hall bar.
\end{abstract}

\pacs{71.10.Pm, 73.63.Nm}
\maketitle

\section{Introduction}
Charge fractionalization, the appearance of quasiparticles which carry a fraction of the charge unit,  is a remarkable effect that is found in certain quantum many-body systems with unusual properties. Well studied examples are the two-dimensional electron fluids of the quantum Hall effect, and ever since the basic theoretical understanding of the fractional effect was established it has been known that the fundamental quasiparticles in these systems are fractionally charged and satisfy fractional statistics \cite{Laughlin83,Halperin84,Arovas84}. Experimental studies of current fluctuations due to charge tunneling have indeed confirmed the presence of fractional charge carriers\cite{Goldman95,Saminadayar97,Picciotto97}.

More recently it has been suggested that excitations with fractional charge may appear in one-dimensional systems described by Luttinger liquid theory \cite{FisherGlazman97,Pham00}. Here the fractionalization is linked to chiral separation of charges that are introduced in the system \cite{Safi95,Safi97}, so that fractions of a unit charge move to the right and the left, respectively. In addition to theoretical works predicting this effect there have been suggestions of experiments that could detect the fractional charges \cite{Trauzettel04,LeHur08,Berg09}, and an experiment has been performed that confirms the expected left-right asymmetry of the current  associated with the injection of charges in the system \cite{Steinberg08}.

There are however important differences between the fractionalization effect  in the two systems, since the quantum Hall fluids are  incompressible whereas a Luttinger liquid is gapless. This difference is of importance both for the question of uniqueness and of sharpness of the fractional charge.
In the quantum Hall case, these properties of the quasiparticles follow from the topological properties of the fluids, while for the Luttinger liquid there seems not to be any unique value associated with the fractional charges, which are instead determined by the way these excitations are created.

In a previous publication, three of us have examined  questions concerning values and sharpness of fractional charges in Luttinger liquids \cite{LeinaasHorsdal09} (see also Ref.~\onlinecite{Leinaas10}). The conclusion is that fractional charges in such systems can be sharp, not in an absolute sense, but in the sense that the charge fluctuations are indistinguishable from the background fluctuations of the ground state. These charges can take different values, depending on initial conditions, and we have in particular examined the difference between the situations where integer charges are introduced suddenly or adiabatically into the Luttinger liquid.

The purpose of the present paper is to follow up this work by studying in some detail the time evolution of charges that are introduced as edge excitations in a quantum Hall bar, where the non-trivial Luttinger liquid behavior is due to interactions between charges on the two edges \cite{Oreg95,Horsdal07}.  We have performed explicit calculations of the time evolution of pulse shapes under transitions between regions with non-interacting and interacting edges, and focus in particular on what happens under (quasi-)adiabatic evolution from integer to fractional charges. 

Our Hall bar geometry  is essentially the same as the one recently studied in Ref.~\onlinecite{Berg09}, where it was suggested that 
the charge fractionalization could be detected by noise measurements. Our analysis supports this claim, but our conclusions differ both on the expected values of the fractional charges, and on the optimal strategy for detecting them.

\section{The effective theory }
\begin{figure}[h]
\begin{center}
\includegraphics[width=8cm]{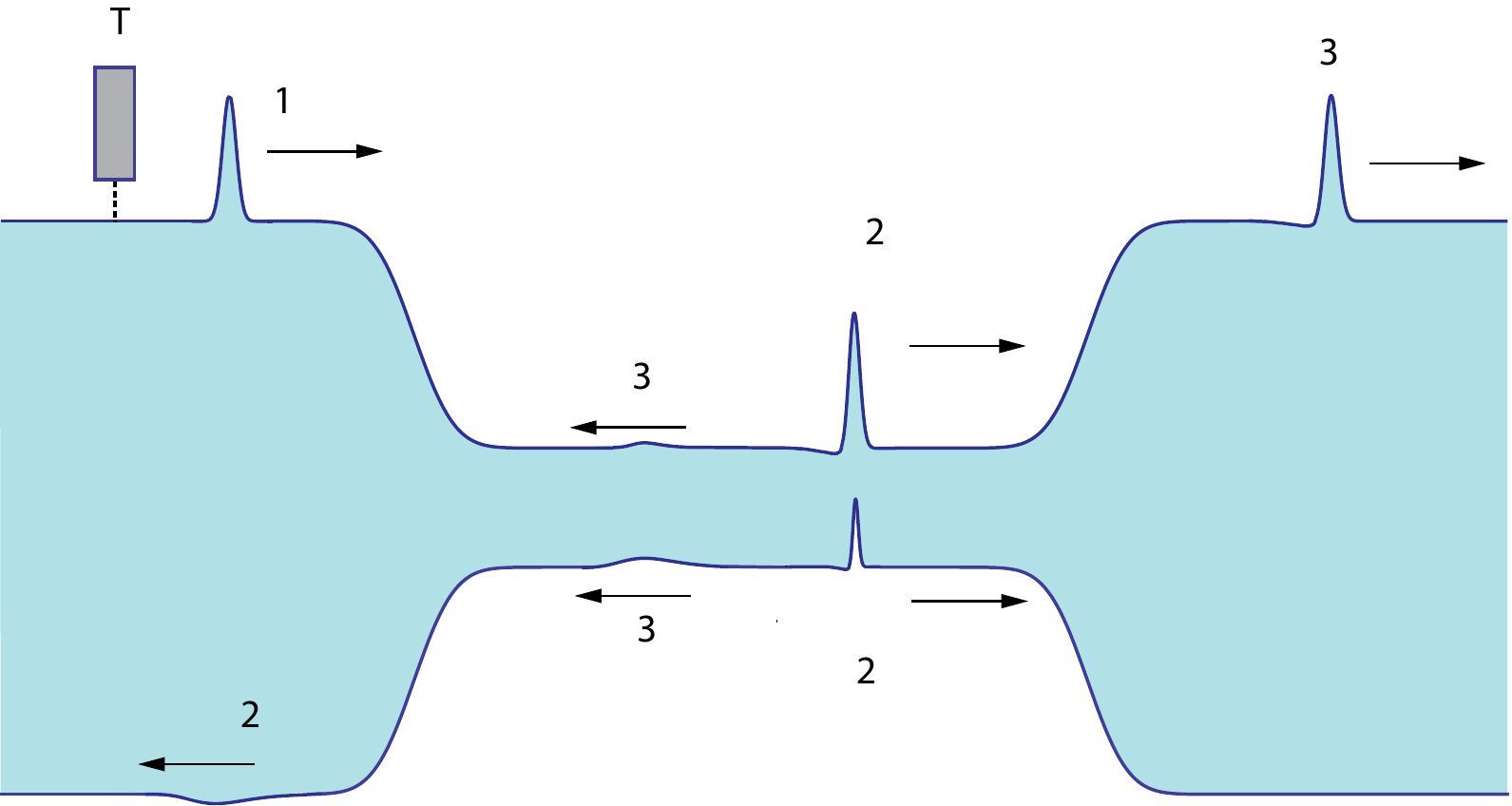}
\end{center}
\caption{\small (Color online) Schematic illustration of the Hall bar with a constriction. At the point $T$ an electron tunnels into the upper edge, shown in the form of sharply defined modulation of the edge (1). The pulse travels to the right, and when entering the constriction, the charge pulse is separated in a reflected and a transmitted pulse (2), with the transmitted pulse appearing in the form of a correlated  modulation of both edges. This pulse is further split into a reflected one, moving within the constriction, and a transmitted one that propagates to the right outside the constriction (3).
 \label{Hallbar}}
\end{figure}

We consider a quantum Hall bar with a constriction at (Landau level) filling fraction $\nu=1$, see Fig.~\ref{Hallbar}.
The electrons are assumed to be fully polarized, so that the spin can be suppressed in the description. In the region of the constriction the two edges are sufficiently close to allow interaction across the sample, but sufficiently far apart to completely suppress charge tunneling between the edges. Outside the constriction, the separation between the edges is much larger, and interaction across the bar is completely suppressed. 

Electrons are assumed to be inserted by tunneling into the system at one of the edges in the region outside the constriction, and to travel with the edge current from the initial region into the constriction.  The figure shows the qualitative picture for the propagation of the electron charge from the point of insertion. 

With no tunneling the charges on each edge are separately conserved and can  be described by two independent fields $\gr_\pm$, which measure the edge charge densities relative to their ground state values. The action can be separated in two parts,
\be{action}
S=S_0+S_{int}
\ee
where $S_0$ describes free spin-polarized electrons, and $S_{int}$ is the electron-electron interaction,
\be{int}
S_{int}=
-{1\over 
{2}}\sum_{\chi=\pm} \int dt \,dx_1 dx_2 \left[\gr_\chi (x_1,t)\,V_1(x_1,x_2)\,\gr_\chi (x_2,t)+
\gr_\chi (x_1,t)\,V_2(x_1,x_2)\,\gr_{-\chi} (x_2,t)\right]
\ee
where $V_1$ is the interaction between charges on the same edge and $V_2$ the interaction between charges on opposite sides. Symmetry between the two edges is assumed, and $x$ is the linear coordinate in the direction of the symmetry axis.
For a soft edge profile, and for sufficiently low energies, we may make the local approximation $V_a(x_1,x_2)=V_a(x_1)\gd(x_1-x_2),\;a=1,2$. The resulting action is that of a Luttinger model 
\be{action2}
S=\pi\hbar \int dxdt\left(\pd_x\Theta\,\pd_t\Phi-\half v(x)\left[{1\over g(x)}(\pd_x\Phi)^2+g(x)(\pd_x\Theta)^2\right]\right)
\ee
with variable parameters $v(x)$ and $g(x)$ given by
\be{effpar}
v(x)&=&\sqrt{\left(u+{1\over{2\pi\hbar}}V_1(x)\right)^2-\left({1\over{2\pi\hbar}}V_2(x)\right)^2}\nn\nn
g(x)&=&\sqrt{\frac{u+{1\over{2\pi\hbar}}(V_1(x)-V_2(x))}{u+{1\over{2\pi\hbar}}(V_1(x)+V_2(x))}} \, ,
\ee
where $u$ is the Fermi velocity of the non-interacting theory. 
The two fields $\Phi$ and $\gQ$ are related to the charge densities through
\be{fields}
\gr_\pm= {\frac{1}{2}} ( \pdn x\Phi\mp\pdn x\gQ )
\ee

The description in terms of the effective parameters $v(x)$ and $g(x)$, as given by  \pref{effpar}, depends on the assumption of a smooth, quasi-adiabatic transition from the wide to the narrow part of the Hall bar. The expressions  are then the same as with $x$-independent interactions $V_1$ and $V_2$ \cite{Haldane81}.  A smooth modulation of the edge profile is desirable for the study of intrinsic physical properties of the system within the constriction, in which case effects that depend on the precise profile of the edges in the transition region are less important. From now on, we shall employ the effective model \pref{action2}.  We have checked numerically that for typical parameters, and pulse shapes, used in the below analysis, this effective model gives essentially the same result as the non-local microscopic theory defined by  \pref{action} and \pref{int}. The latter is presumably more accurate, but also computationally much more demanding.

\section{Transition and reflection of charged pulses}
\subsection{Multiple reflection analysis}
It is convenient to change to new, $g$-dependent, charge density variables, defined by
\be{dens}
f_\pm={\frac{1}{2}} (\pd_x\Phi\mp g\pd_x\Theta)
\ee
They satisfy the following field equations,
\be{fieldeq}
\pd_t f_\pm=\mp\pd_x(v f_\pm)\pm\half v\frac{\pd_x g}{g}(f_++f_-)
\ee
as can be derived from the action \pref{action2}.
In regions where $g$ is constant $f_\pm$ define the two chiral components of the charge density, corresponding to the right and left moving parts of the edge fields. For the case $g=1$ this separation in terms of the right and left moving components is identical to the separation of the total charge into components of the two edges, so that $f_\pm=\gr_\pm$. However, when $g\neq 1$ there is a difference between these two ways to decompose the total charge, and we have 
\be{decomp}
f_\pm=\hf(1\pm g)\gr_++\hf(1\mp g)\gr_-
\ee
As a consequence there is for a purely right(left) moving mode $(f_\mp=0)$ a unique ratio between the charge densities of the two edges
\be{ratio}
\gr_\pm/\gr_\mp=\frac{g+1}{g-1}\,,\quad (f_\mp =0)
\ee
In a region with constant $g$ a right(left) moving charge will thus decompose in two parts, with a charge distribution on the upper(lower) edge and a co-moving mirror image of the charge distribution on the lower(upper) edge, and with a fixed ratio $(g+1)/(g-1)$ between the two charges \cite{LeinaasHorsdal09}.

The Luttinger model for the quantum Hall bar gives rigid constraints for charge transfer between two regions with different values of the interaction parameter $g$. This follows since, on one hand, the charge is separately conserved for each edge of the Hall bar, and on the other hand, the charge with a given chirality splits up in components on the two edges, which have a unique ratio determined by the value of $g$. As a result the transmission coefficient $T$ and the reflection coefficient $R$ for scattering of a charge on the boundary between two regions with different values of the interaction parameter are determined as
\be{transit}
T= \frac{2g}{g'+g}\,,\quad R=\frac{g'-g}{g'+g}
\ee
with $g'$ as the value of the interaction parameter in the region of the incoming charge and $g$ as the parameter in the region of the transmitted charge. In particular, for scattering from the non-interacting region ($g'=1$) into the interacting region ($g\neq 1$) we have $T=2g/(1+g)$ and $R=(1-g)/(1+g)$, consistent with the results  of Ref.~\onlinecite{Safi95}. One should note that the values  \pref{transit} of $T$ and $R$ are independent of the shape of the scattered charge density and of the functional form of $g(x)$ in the region of interpolation between the two values $g'$ and $g$ of the interaction parameter.

\begin{figure}[h]
\begin{center}
\includegraphics[width=14cm]{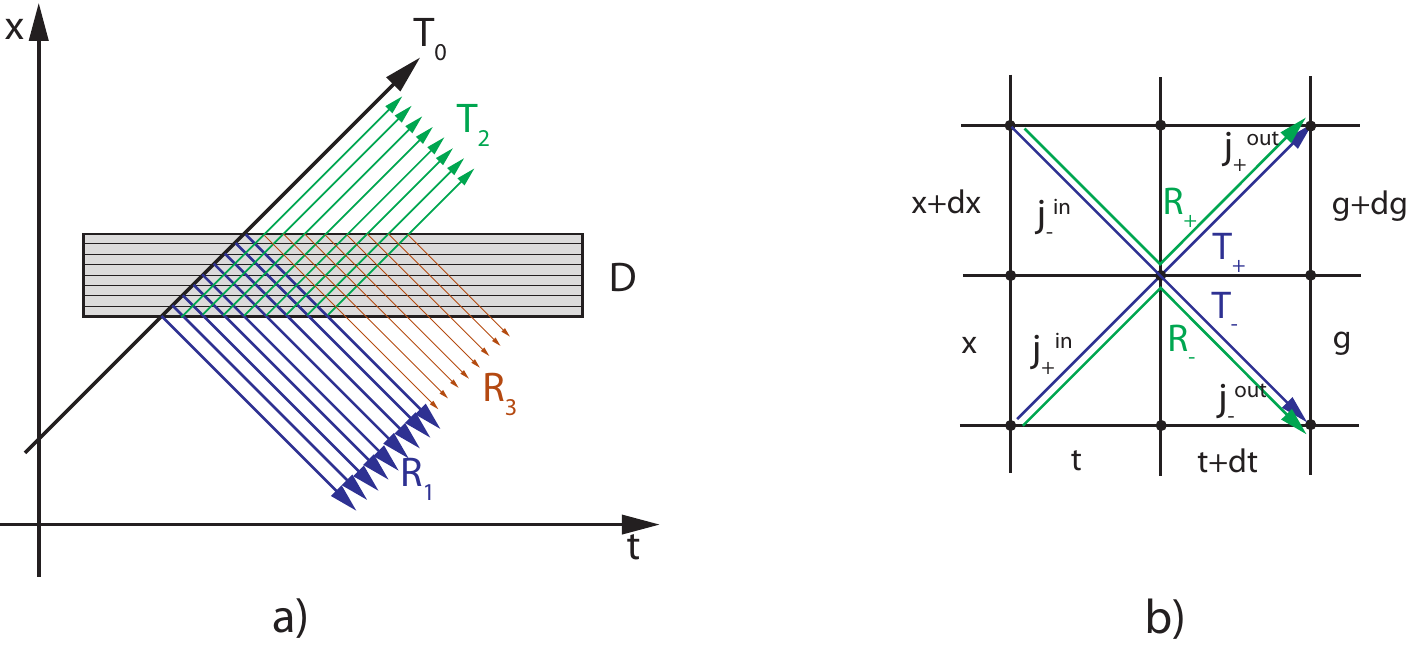}
\end{center}
\caption{\small (Color online) Schematic  illustration of charge reflections and transmissions in the discretized model. In a) a sharply defined charge pulse, represented by the thick black line enters a region $D$ with a variable value of the interaction parameter $g$.  On each step, where $g$ changes, it is split into a transmitted and a reflected charge. The total transmitted and reflected charges can be separated in charge components, characterized by the number of internal reflections they have experienced during the transit through $D$.  The transmitted component $T_0$, with no internal reflections, corresponds to the sharply defined front pulse of the transmitted charge. The reflected component $R_1$, with one internal reflection, is broadened during the transit of $D$. It is clear from the illustration that the width of each charge component increases with the number of reflections. In b) the transmission and reflection at a single step is shown in a space-time diagram. The rules for reflection and transmission \pref{transit}, when applied at  each step, gives rise in the continuum limit to the effective field equation of the system, as discussed in the text. \label{Reflection}}
\end{figure}
It is instructive to consider a discretized model, as shown schematically in Fig.~\ref{Reflection}, for the transmission and reflection of charges in a region where $g$ changes. The value of $g$ (and of $v$) is then assumed to change in a stepwise fashion, with transmission and reflection of charge taking place at each step. The scattering at individual steps (as illustrated in Fig.~\ref{Reflection}b)) is assumed to satisfy the relations \pref{transit} for the (local) transmission and reflection coefficients.

The outgoing, transmitted and reflected charges can be decomposed into parts defined by a given number of reflections inside the interval where $g$ changes, and an interesting point is that in the continuum limit of the discretrized function $g(x)$, each of these charge components tends to a finite value, which takes a simple form when expressed in terms of the interaction parameter.
 With $g=1$ in the region of the incoming pulse and $g\equiv\exp(-2\Gamma)$ in the region of the transmitted pulse, the expansions in number of reflections are expressed as 
\be{TRexp}
&&T=  \sqrt g\,  (1-\half \Gamma^2+{5\over 24} \Gamma^4-{61\over720} \Gamma^6 +{277\over8064}\Gamma^8 +{\cal O}(\Gamma^{10}))\nn
&&R=\Gamma-{1\over 3} \Gamma^3+{2\over 15}\Gamma^5 -{17\over 315}\Gamma^7+{62\over 2835}\Gamma^9 +{\cal O}(\Gamma^{11})
\ee
where the number of reflections corresponds to the power of $\Gamma$. The series sum up to $T=\frac {2g}{1+g}$ and $R=\frac{1-g}{1+g}$, consistent with the asymptotic expressions given in \pref{transit}.

It is of interest to note that the effective theory described by the action \pref{action2} can be viewed as a {\em local} implementation of the relations \pref{transit} for transmission and reflection of charges. Thus we expect the continuum limit of the discrete  model outlined above to give a faithful representation of the effective theory \pref{action2}. To explicitly show this we consider the scattering of charge on a single step, as illustrated in Fig.~\ref{Reflection}b, in the form of a matrix equation,
\be{scattering}
\left(\matrix {j_+^{out}\cr j_-^{out}}\right)=\left(\matrix{T_+&-R_+\cr -R_-&T_-} \right)
\left(\matrix {j_+^{in}\cr j_-^{in}}\right)
\ee
where $j_\pm=\pm v f_\pm$ define the charge currents. For an infinitesimal change in $g$, we have according to \pref{transit}, $T_\pm=1\pm dg/2g$ and $R_\pm=\pm dg/2g$, and for corresponding infinitesimal changes in the coordinates $x$ and $t$ between the {\em in} and {\em out} states, the scattering equations can be written as (see Fig.~\ref{Reflection}b)),
\be{scattering2}
j_+(x+dx,t+dt)=(1+{dg\over{2g}})\,j_+(x,t)-{dg\over{2g}}\,j_-(x+dx,t)\nn
j_-(x,t+dt)=(1-{dg\over{2g}})\,j_-(x+dx,t)+{dg\over{2g}}\,j_+(x,t)
\ee
To first order in the differentials, and by use of the relation $dx=vdt$, the above equations can be re-written as the following differential equation
\be{fieldeq2}
\pd_t j_\pm=\mp v \pd_x  j_\pm + \half v\frac{\pd_x g}{g}(j_+-j_-)
\ee
and it is straightforward to check that it is equivalent to the field equation \pref{fieldeq}.

An important point to note from the discretized model, clearly demonstrated in Fig.~\ref{Reflection}, is that the transmitted pulse has a distinct separation in a front pulse and a tail. Thus, with a sharply defined incoming pulse, as in Fig.~\ref{Reflection}, the leading transmitted pulse, which has no internal reflections, will be  equally sharp. This is different from the other transmitted components, which are broadened by the reflections. With $L$ as the width of the scattering region the contribution from $n$ reflections has a width that is in fact proportional to $nL$, since each additional pair of backward-forward reflections will spread any part of the pulse over the width $L$,  as is illustrated in Fig.~\ref{Reflection}. The net result is that the transmitted pulse obtains a long tail in addition to the sharply defined front pulse. This effect is present also in the continuum limit. We note that the transmitted front pulse has a shape that is insensitive to the functional form of  $g(x)$ and $v(x)$, while the precise form of the tail will depend on these functions.

\begin{figure}[h]
\begin{center}
\includegraphics[width=16cm]{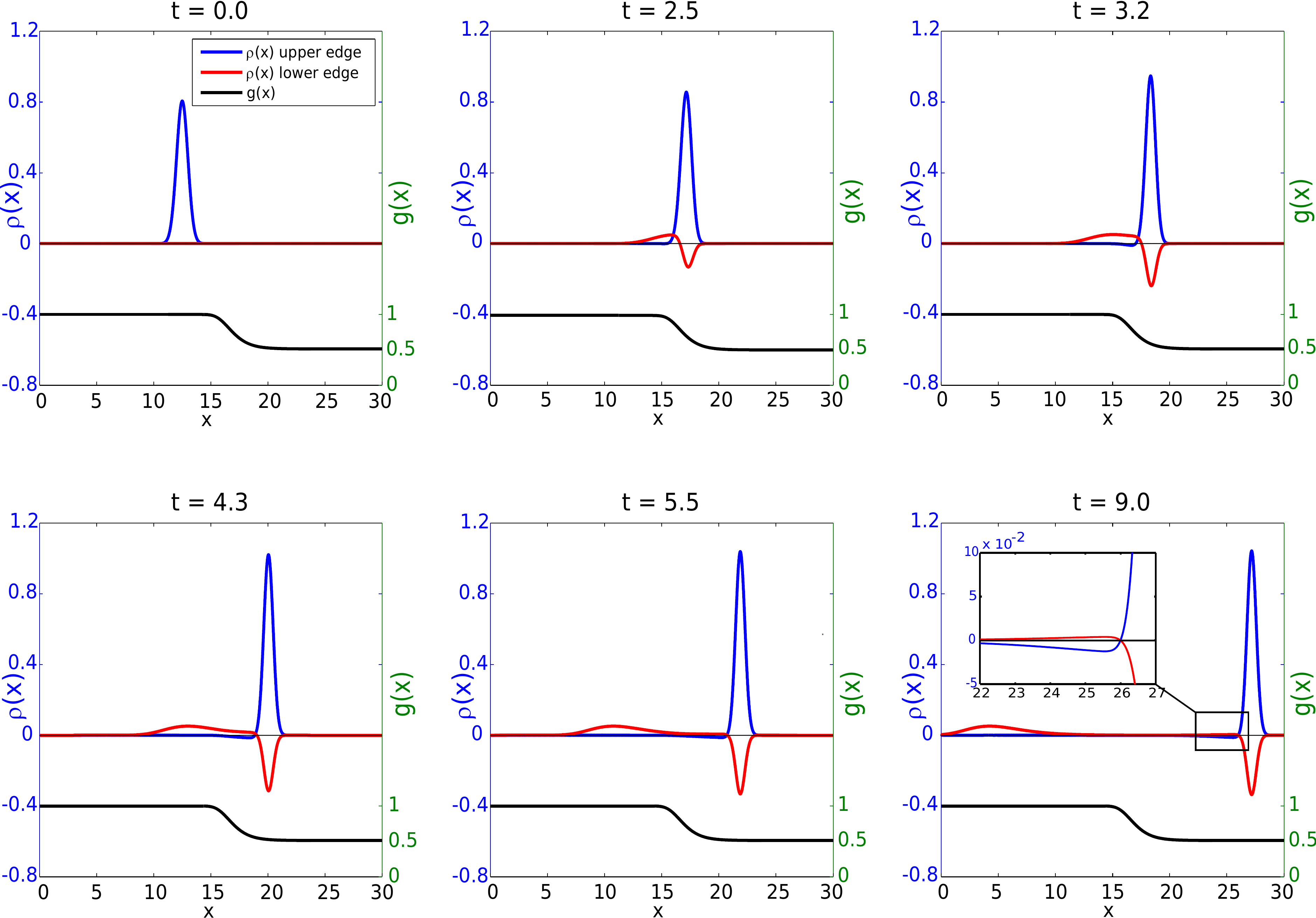}
\end{center}
\caption{\small (Color online) Numerical simulation of the scattering of an incoming pulse due
to changes in the interaction parameter from $g=1$ to $g=0.51$. The figure displays the charge density at the upper (blue line) and the lower edge (red line) at different times. The lower (black) curve represents the changing value of the parameter $g(x)$. Initially a pulse at the upper edge travels to the right in a region where $g=1$. As it {progagates through the transition region} there are two distinct charges appearing on the lower edge. The mirror charge has the same shape as the pulse on the upper edge, but a smaller value and the opposite sign. The reflected
part is much broader and its width depends on the length of the transition
region. A close examination (see inset) reveals that the
pulses traveling to the right divide into a sharp front pulse and a long tail
arising from multiple reflections within the transition region. The shape of the tail and of the reflected pulse is
determined by the profile of $g(x)$ and $v(x)$. The front pulse is independent
of the shape of the functions, though its total charge is determined by the
final value of $g$ and its width on the final value of $v$. Since $v$ decreases the
front pulse gets compressed. \label{Timestep}}
\end{figure}

The distinction between the front pulse and the tail is directly related to the discussion given in Ref.~\onlinecite{LeinaasHorsdal09} of how a charge that is adiabatically introduced into the system is separated into a local charge and a non-local charge that is evenly distributed over the system. The charge of the front pulse, $T_0 =\sqrt g$, is indeed identical to the charge found in Ref.~\onlinecite{LeinaasHorsdal09} for the local part of charge. The tail of the charge distribution, defined by multiple reflections in the transition region, will tend to the evenly distributed non-local charge in the adiabatic limit. For the reflected charge there is no component corresponding to the sharply defined front pulse, since all components are broadened by the reflections. For a smooth function $g(x)$ the reflected pulse will therefore give a week, broad signal. 

\subsection{Numerical simulations of pulse shapes}

Here and in the following we shall present results of numerical solutions of the field equation \pref{fieldeq} with a discretized time coordinate, and with gaussian initial pulses at the point of tunneling. In this, and subsequent,  simulations the  shape of the edges are expressed in terms of error functions as
\be{edgeprof}
y(x)= \pm\hf \left\{{ \frac{W-W_c}{2} }\left[{\rm erf}\left({{x-\ell/2}\over b/4}\right)-{\rm erf}\left({{x+\ell/2}\over b/4}\right) \right]+{W} \right\}
\ee
with $x$ the coordinate along the Hall bar and $y$ the transverse coordinate, with $W$ as the transverse width of the bar outside the constriction and $W_c$ as the width within the constriction. The parameter $b$ defines the longitudinal width of the transition region and $\ell$ the length of the constriction.
{In dimensionless units the parameter values used in the evaluations are $W=10$, $W_c=0.1$ and $b=80$. Below we shall consider two situations. In the first, where the tunneling occur at an external lead, we set  $\ell=120$, and in the second, where the tunneling is  within the constriction, we set $\ell =180$. All other parameters are taken equal.} Furthermore the width of the initial gaussian pulse of the tunneling charge is chosen as $\gD=0.6$. A screened interaction between the charges on the two edges is assumed, and is here modelled by a Gaussian potential with damping length $d \approx 0.8$.  {The strength of the interaction is chosen to give $g=0.64$ for the interaction parameter and $v_c=21$ for the effective velocity within the constriction.}

{ 
All the above parameters are dimensionless and we need a suitable timescale for later evaluation of current and noise. 
Our choice is the propagation time between the endpoints of the constiction, \textit{i.e.}, the length of
the constriction divided by $v_c$.
The Luttinger parameter $g(x)$ will not change until the distance between edges are of the same size as the interaction length. Thus the natural length of the 
constriction is not $\ell$, but an effective length $\ell_{\mathrm{eff}}$ which we define to be the distance between the points where $g(x)$ is halfway between
1 and and the value of $g$ in the constriction. We will scale time with $t_0 = \ell_{\mathrm{eff}}/v_c$ and frequency with $\omega_0 = 2\pi v_c/\ell_{\mathrm{eff}}$ in
current and noise plots. }

In  Fig.~\ref{Timestep}, we simulate the {time evolution of a charged pulse incident on a single step in the width corresponding to a transition from a region with $g=1$ at the far left to a region with  $g=0.51$ at the far right. The initial pulse} is in the transition region split into a right moving, transmitted pulse and a left moving reflected one. The figure clearly displays the front pulse of the transmitted charge, which is followed by a long, weak tail. Also the the broad reflected pulse is clearly visible. The two components of the front pulse, the main component on the upper edge and the mirror image on the lower edge, are both shown in the figure. 

\begin{figure}[h]
\begin{center}
\subfigure[]{\label{FinalPulse}\includegraphics[scale=0.5]{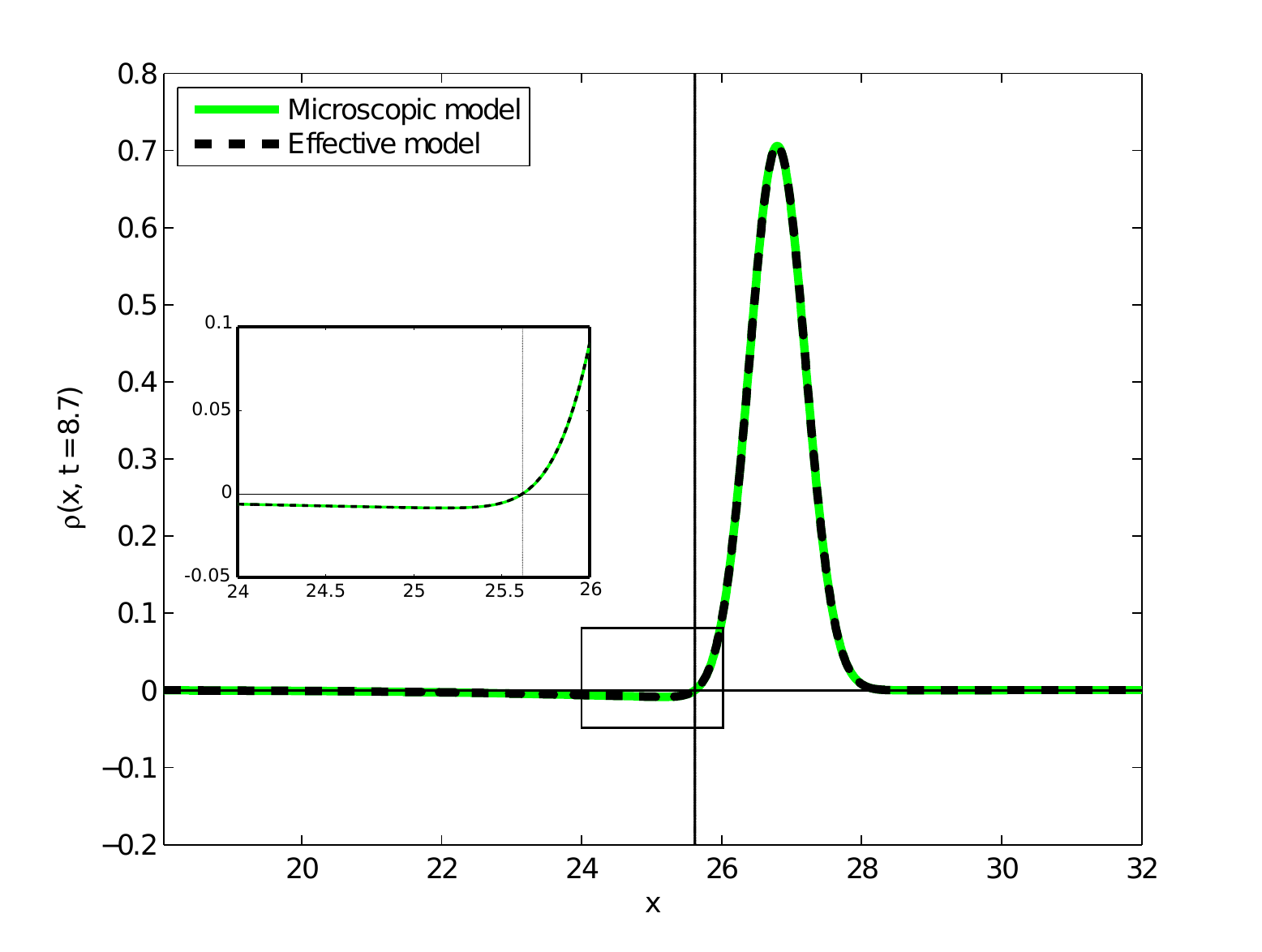} }
\subfigure[]{\label{GaussFit}\includegraphics[scale=0.5]{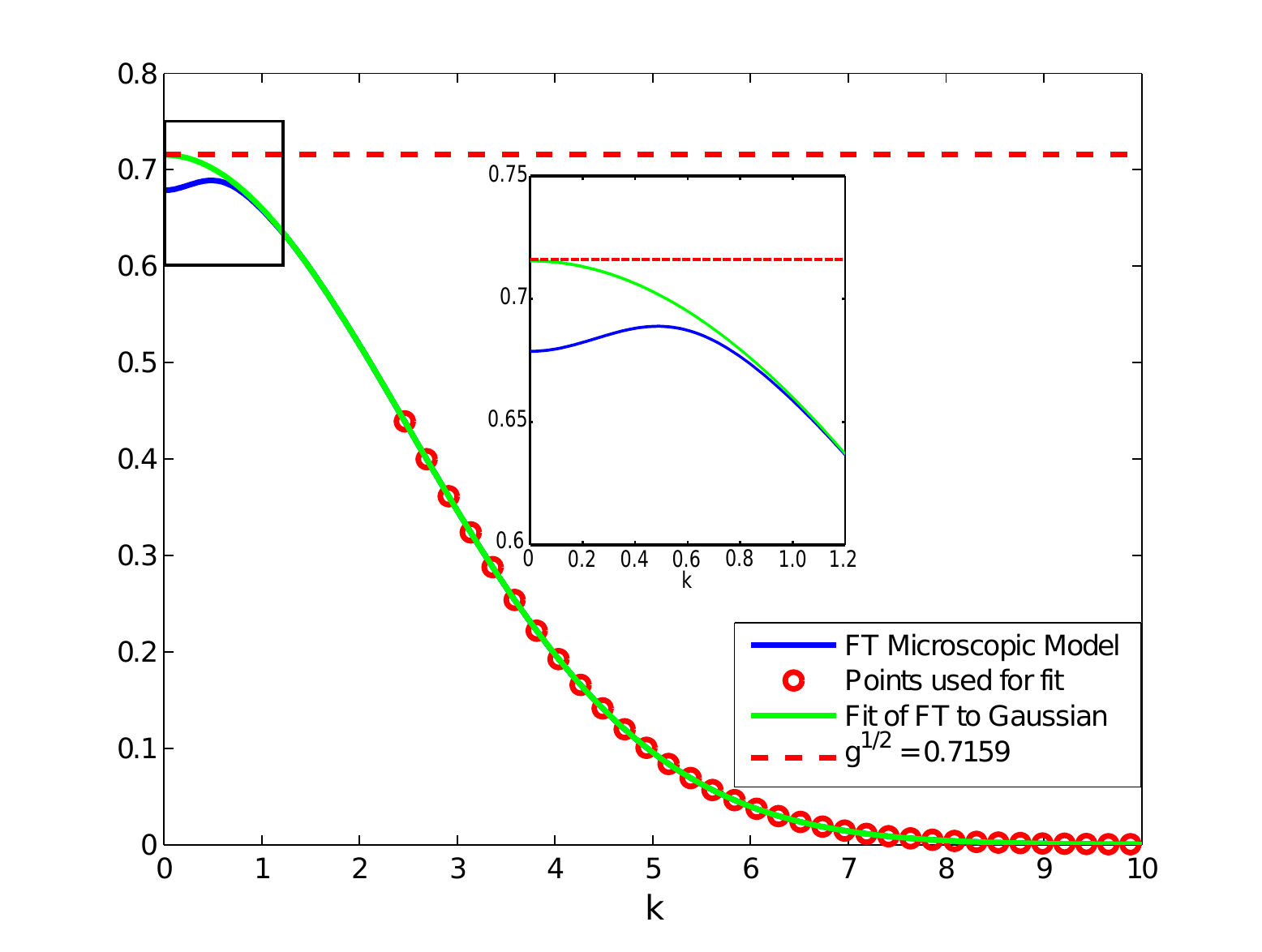} }
\end{center}
\caption{\small (Color online) Estimation of the charge of the front pulse. (a) Sum of the density distributions at the upper and lower edge at the final time step of Fig.~\ref{Timestep}. Only the rightmost part of the density is included, which corresponds to the rightmoving charge. The inset helps to distinguish between the front pulse and the tail. (b) Fourier transform of the density distribution of the microscopic model (blue curve), shown with a Gaussian fit to the curve (in green). The red rings are points used to determine the fit parmeters (only every 20th point is shown). From the inset we clearly see that the blue line approaches the value of the total transmitted charge for $k\rightarrow 0$, while the green line approaches $\sqrt{g}$.}
\end{figure}

We have applied two different mehods to estimate the charge of the front pulse from the numerics. The first method is to integrate the density distribution only in the region corresponding to the front pulse. Fig.~\ref{FinalPulse} shows the sum of the density distributions at the upper and lower edge at the final time step of Fig.~\ref{Timestep}. We can identify the front pulse as the part of the density distribution where the density is positive, that is, for $x>25.6$. A numerical integration in this region gives the value $0.7029$ for the effective model and $0.7028$ for the microscopic model, as compared to $\sqrt{g}=\sqrt{0.5125}=0.7159$. 
It is interesting to compare these results to the total transmitted charge, which is given by the transmission coefficient, $T=\frac{2g}{1+g}=0.6777$.
The second method is to estimate the charge from a Gaussian fit to the front pulse. The blue line in Fig.~\ref{GaussFit} shows the Fourier transform of the density distribution from the microscopic model in Fig.~\ref{FinalPulse}. We see that for $k\rightarrow 0$ the Fourier transform approaches the total transmitted charge, $0.6777$. The frontpulse is assumed to retain its shape through the transition region, we therefore fit a Gaussian to the Fourier transform for $k>k_{co}$, where $k_{co}$ is a cut off. With the cut off equal to the width of the Fourier transform of the Gaussian we find that the charge is given by $0.7154$. When varying the $k_{co}$ from $0.6$ to $1.4$ of the width of the Fourier transform of the Gaussian, the estimated charge varies from $0.7153$ to $0.7162$. This way of estimating the charge gives a better agreement with the predicted value of  $ \sqrt g$, than the first method. This is not surprising since, with the parameter values we use, the step and pulse width are comparable in size. This means that the positive region is likely to have contributions from higher order reflections, and these will be included in the first estimate. In the second method only the high momentum components are used to fit the gaussian profile, and these are not expected to be sensitive to the finite width of the step.

\section{Current fluctuations and fractional charges}
By measuring current fluctuations, one can gain information about the presence of fractional charges in the system. {Ref.  \onlinecite{Trauzettel04}  has considered scattering of charges on an impurity in a system with an effective Luttinger parameter $g$, where noise is produced in the backscattered current.} The expected value for the scattered charge, derived from the $\gw\to 0$ limit of the Fourier transformed noise,  was found to be $g$ times the electron charge. 
In Ref. \onlinecite{Berg09}, which considered the Hall bar geometry discussed above, it was suggested that the fractional charge $(1-g)e/(1+g) $, given by the reflection coefficient in \pref{transit}, could be observed by measuring the tunneling noise in the reflected current on the lower edge (see Fig.~\ref{Hallbar}).
However, secondary pulses created from reflections on the second boundary will in this case neutralize the charge to form an extended reflected signal with zero total charge, and the information about the presence of fractional charge therefore has to be extracted from the 
$\gw\neq 0$ part of the noise. 

We too consider  the current noise due to tunneling, but focus on the above distinction between the front pulse, and the broadened tail, of a charge transmitted through the constriction. The front, which is the robust part of the pulse, carries  the charge $\sqrt g \, e$, and we take this as the natural definition of a fractional charge value within the constriction. The expressions for the noise are found by a simple, semiclassical approach, and they essentially agree with the expressions given in Ref.~\onlinecite{Berg09}. Although it might be more difficult to measure, we shall consider the noise also in the current inside the constriction, since it provides a clear signature for fractional charge. In the last section we shall propose an alternative tunneling geometry where there is  a clear signature in a  more readily measurable current. 

\subsection{Noise in tunneling currents}
The tunneling is described as a Poisson distributed sequence of events, each characterized by a sharply defined current pulse $I_s(t)$, with $\int  I_s(t) dt = e$. The total current at time $t$, restricted to a (large) time interval $T$, at a point close the to the point of tunneling, is a distribution of $N$ pulses superimposed on the background edge current, $I_{tot}=I_{tun}+I_{0}$, with
\be{seq}
I_{tun}(t)=\sum_{n=1}^N I_s(t-t_n)
\ee
where $t_n$ denotes the instant of a tunneling. The tunneling events are assumed to occur randomly within the given time interval, thus giving  a constant average current
\be{ave}
\mean{I_{tun}}=\sum_{n=1}^N {1\over T}\int_0^T dt_n  I_s(t-t_n)=e{N\over T} \, .
\ee
This average is held fixed, and in the following the limit $T,N\to\infty$ is taken whenever convenient.

The noise of the current is defined as the Fourier transform of the current-current correlation function
\be{S}
S(\gw)=\int_{-\infty}^\infty dt\, e^{i\gw t}[\mean{I(t)I(0)}-\mean{I(t)}\mean{I(0)}]+c.c. \, .
\ee
With no correlation between the tunneling and background currents we have $S(\gw)=S_{tun}(\gw)+S_{0}(\gw)$, and we focus primarily on the fluctuations in the tunneling current. The assumption of randomly distributed tunneling events implies that the expectation values can be found by independently integrating over the tunneling times $t_n$,
\be{Stun}
S_{tun}(\gw)&=&\lim_{T\to\infty} {\bigg[} \sum_{n}{1\over T}\int_{-\infty}^ {\infty}dt \int_0^T dt_n 
e^{i\gw t}I_s (t-t_n) I_s (-t_n)\nn
&+&\sum_{n\neq n'}{1\over T^2}\int_{-\infty}^{\infty}dt \int_0^T dt_n \int_0^T dt_{n'} e^{i\gw t}I_s (t-t_n) I_s (-t_{n'})\nn
&-&\int_{-\infty}^{\infty}dt e^{i\gw t} \mean{I_{tun}}^2 {\bigg]} +c.c.\nn
&=& 2{N\over T}\tilde I_s(\gw)\tilde I_s(-\gw) 
= 2e |f_s(\gw)|^2 \mean{I_{tun}}
\ee
where $\tilde I_s(\gw)$ is the Fourier transform of $I_s(t)$ and $f_s(\gw)$ is the normalized profile function, defined by $\tilde I_s(\gw)=e f_s(\gw)$. This definition gives $f_s(0)=1$ and therefore $S_{tun}(\gw\to 0)=2e  \mean{I_{tun}}$, consistent with the expected noise-current relation in the non-interacting region ($g=1$). As shown by the above expression, the noise can be viewed as a single-pulse effect, since it is determined by the Fourier transform $\tilde I_s(\gw)$ of the charge pulses generated by the tunneling. 

\subsection{Noise in the reflected current}
Due to the linear propagation of the charged pulses through the Hall bar, the current fluctuations at other points along the bar are related in a simple way to the fluctuations at the initial point. {Note that no additional noise is produced by the constriction itself due to the adiabatic transition from the outside to the inside.} We consider first the fluctuations in the reflected current on the lower edge (Fig.~\ref{Hallbar}). The relation between the current at this point and at the initial point can be written as
\be{prop}
I_R(t)={\int_{-\infty}^\infty} dt' G_R(t-t')I_{tun}(t')
\ee
with $G_R(t)$ as the propagator between the two points. The Fourier transform takes the form
\be{propfour}
\tilde I_R(\gw)=R(\gw)\tilde I_{tun}(\gw)
\ee
where $R(\gw)=\int dt \exp(i\gw t) G_R(t)$ is the frequency dependent reflection coefficient, with $R(0)$ as the full reflection coefficient. It follows from the definition of $S(\gw)$ that the noise at the point where the reflected current is measured is given by
\be{refnoise}
S_R(\gw)&=&|R(\gw)|^2S_{tun}(\gw)\nn
&=&2e|R(\gw)|^2 |f_s(\gw)|^2 \mean{I_{tun}}
\ee
The mean reflected current is $\mean{I_R}=R(0)\mean {I_{tun}}$. However, the transmission coefficient for the Hall bar across the constriction is $T=1$, since we have the same value $g=1$ on both sides \cite{Safi95}. This means that the reflection coefficient is $R\equiv R(0)=0$ and therefore there is no contribution to the average edge current from the tunneling, $\mean{I_R}=0$. Also the tunneling contribution to the noise vanishes in the $\gw\to 0$ limit.

\begin{figure}[h]
\begin{center}
\includegraphics[width=9cm]{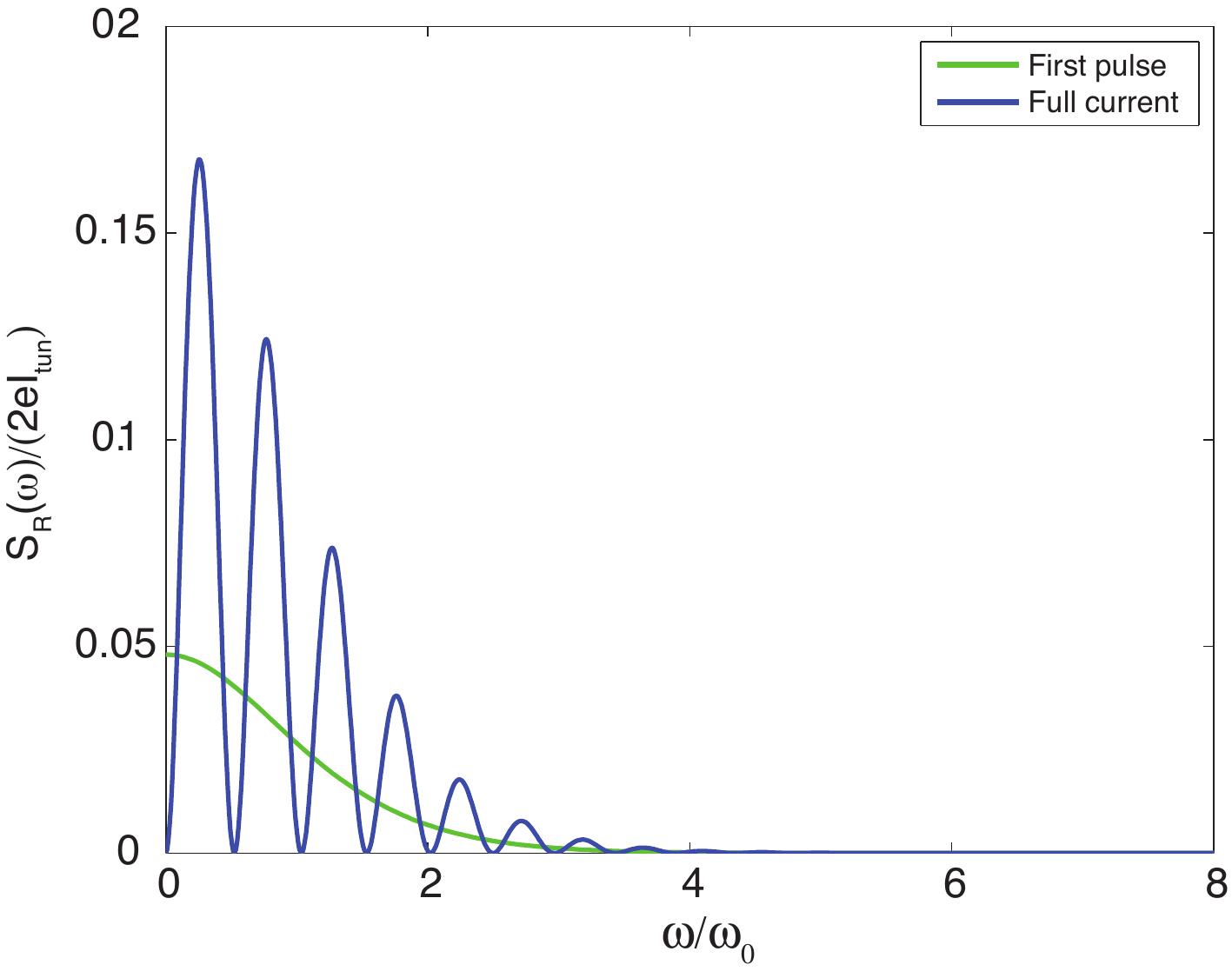}
\end{center}
\caption{\small  (Color online) The figure displays the result from a numerical evaluation of the tunneling contribution to the noise, $S_R(\gw)$, of the reflected current on the lower left edge of the Hall bar. {The value of the interaction parameter is $g = 0.64$ and the frequency scale $\omega_0=2\pi v_c/\ell_{\mathrm{eff}}$
is given by the effective velocity $v_c=21$ and length $\ell_{\mathrm{eff}}=77$ of the constriction.} 
The green curve is the contribution from the first reflected pulse only, which is reflected at the
left boundary of the constriction. The numerically determined value for the $\gw\to 0$ limit of this contribution fits with high precision the theoretical value 
{$[(g-1)/(g+1)]^2\approx 0.048$}. The decay length of the curve is inversely proportional to the width of the reflected pulse, and thereby to the width of the transition region where $g(x)$ changes. The blue curve represents the full noise function, where contributions from the secondary pulses are included. These are pulses that arise from multiple reflections inside the constriction. The value that is found for the full noise function in the limit $\gw\to 0$ is $0$ to high precission, consistent with the expectation that charges of the reflected pulses add up to zero. The oscillations in the blue curve are due to the time shift between pulses reflected at the two boundaries, and the regular form of the curve  is due to the symmetric form of the constriction.   \label{LowerLeft}}
\end{figure}

The vanishing of the average current can be viewed as due to cancellations between charge components associated with repeated reflections between the boundaries of the constriction. In Ref.~\onlinecite{Berg09} it has been suggested that one can nevertheless extract information from the noise, by exploiting the time delay of the secondary reflected pulse, either by filtering out pulses from the second boundary or by measuring the noise function $S_R(\gw)$ for non-vanishing frequencies $\gw$. {The  noise function $S_R(\gw)$ is here obtained by a numerical Fourier transform of the time dependent, reflected current, and is measured relative to $2e\mean{I_{tun}}$.}
The result is displayed in Fig.~\ref{LowerLeft}, which shows both the full function, and the function defined by  including only  the contributions from the first reflected pulse. As expected, the full function tends to $0$ when $\gw\to 0$, and it shows an oscillatory behavior due the interference between contributions from repeated reflections. The decay length of the curve is inversely proportional to the time width of the first reflected pulse, which is in turn proportional to the width of the transition region where $g$ changes. Even though the oscillatory function contains some information about the contribution from the first reflected pulse, shown as the smooth green curve in the figure, and thereby about the charge carried by this part of the reflected signal, to extract a good numerical value from a corresponding experimental curve may seem difficult. Note also that the current and noise shown in the plot in reality sits on the top of the background edge current, which has both non-vanishing  average value and noise.

\subsection{Current noise within the constriction}
\begin{figure}[h]
\begin{center}
\includegraphics[width=9cm]{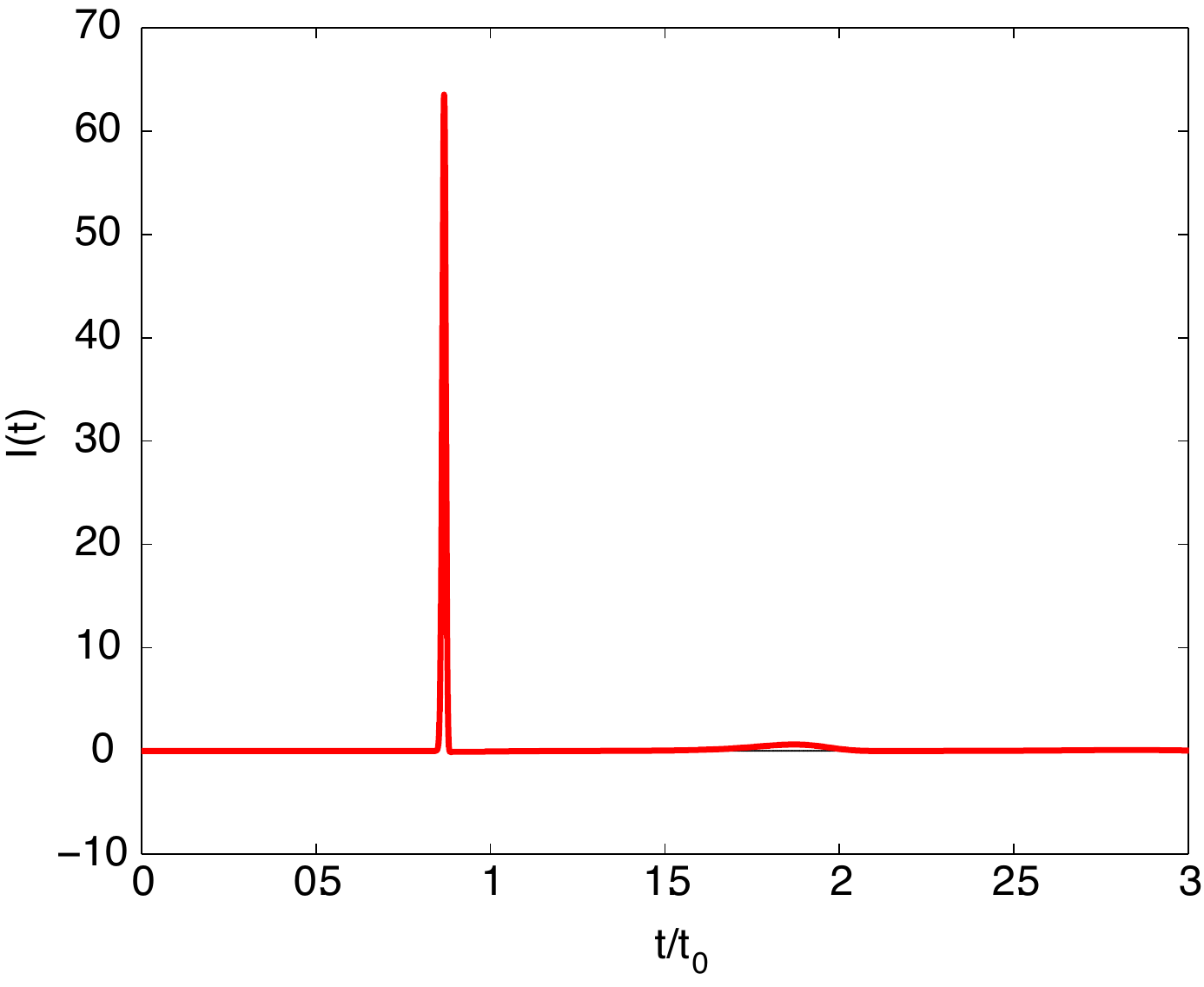}
\end{center}
\caption{\small (Color online) Time dependence of the current due to a single tunneling event, numerically evaluated at a point within the constriction. 
{The time scale $t_0=\ell_{\mathrm{eff}}/v_c$ is given by the effective velocity $v_c=21$ and length $\ell_{\mathrm{eff}}=77$ of the constriction.} The front pulse is seen as the sharply defined peak, followed by the tail which is so small that it is barely visible. The smaller pulse at { $t \approx 1.8$} carries the charge reflected at the right end of the system (pulse 3 in Figure 1). The charge of the pulse is negative, but it gives a positive contribution to the current since it is propagating to the left. It is considerably broader than the first pulse as a consequence of the reflection. At later times more pulses will pass, but their values decrease and their widths increase with each reflection, making them rapidly less significant. There is also a small second reflected pulse, at {$t \approx 2.8$}, that is not visible in the plot. \label{CurrentMiddle}}
\end{figure}

As pointed out, measurements on the reflected current is hampered by  the signal being weak, and broadened by the width of the transition regions where $g$ changes. In fact, in the adiabatic limit the reflected pulses completely disappear into background current. For  the transmitted current, where the front of the pulse is not broadened in the transition region, the situation is qualitatively different. We consider therefore next the current and current noise within the constriction, assuming simply that these can also be subject to measurements. Since the total current, rather than the currents on the separate edges will be more accessible to measurements, we will focus on this quantity, see Fig. \ref{CurrentMiddle}.

The current in the constriction is related to the initial tunneling current in precisely the same way as the reflected current. We write it as
\be{propT}
I_T(t)={\int_{-\infty}^\infty} dt' G_T(t-t')I_{tun}(t')
\ee
with $G_T(t-t')$ as the propagator from the initial point to the point where the current is measured. We similarly have for the Fourier transform
\be{propfour2}
\tilde I_T(\gw)=T(\gw)\tilde I_{tun}(\gw)\,,\quad T(\gw)={\int_{-\infty}^\infty} dt \, e^{i\gw t} G_T(t) \, ,
\ee
and for the current noise
\be{refnoise2}
S_T(\gw)&=&|T(\gw)|^2S_{tun}(\gw) 
=2e|T(\gw)|^2 |f_s(\gw)|^2 \mean{I_{tun}} \, .
\ee
Since the transmission coefficient is unity, $T=T(0)=1$, the average current is simply ${\mean{I_T}}=\mean{I_{tun}}$. Therefore a naive application of the ratio between the noise and average current in the $\gw\to 0$ limit would give $e$ as the charge, rather than a fractional value.

\begin{figure}[h]
\begin{center}
\includegraphics[width=16cm]{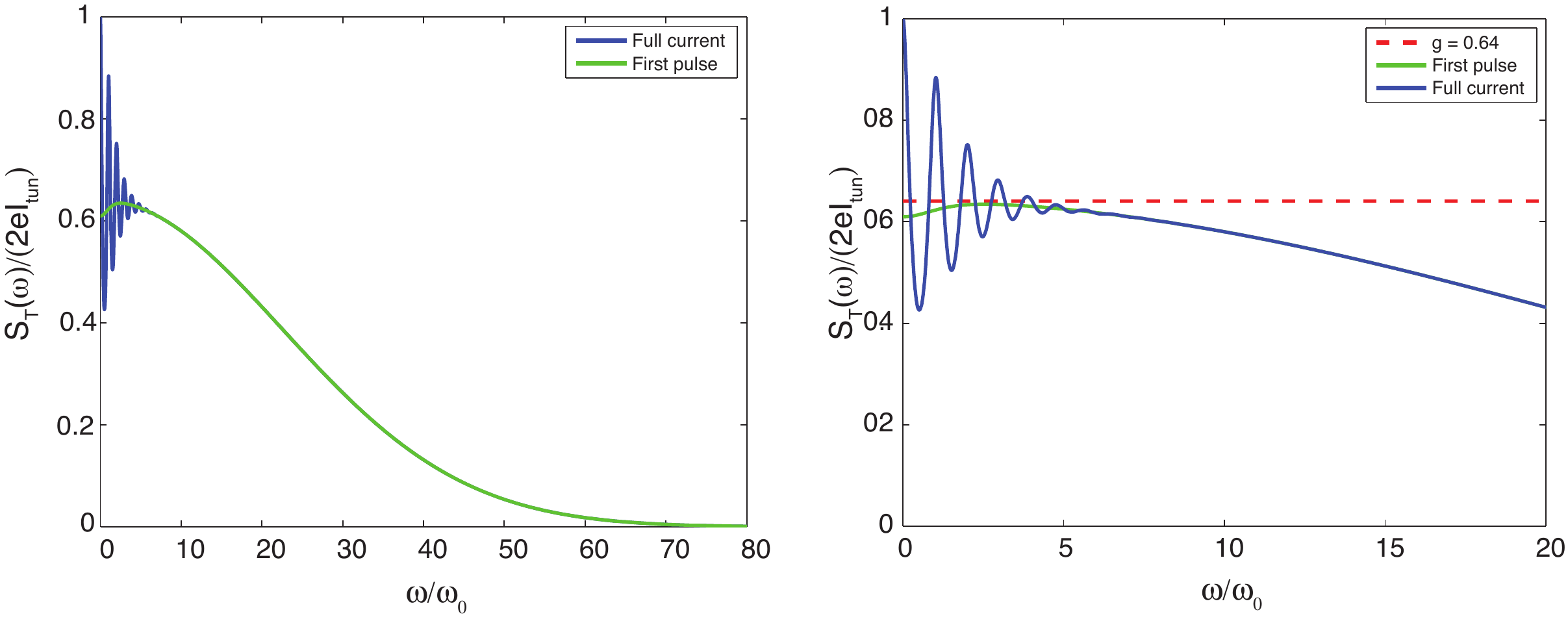}
\end{center}
\caption{\small (Color online) The numerically evaluated noise function $S_T(\gw)$ at a point within the constriction, plotted as a function of frequency $\gw$. The plot to the right shows details of the left plot for small $\gw$. The green curve represents the result when only the first pulse traveling through the constriction is included, while the blue curve gives the result for the full current.  At high frequencies there is overlap between the green and the blue curves, indicating that the high frequency part of the noise function is governed by the front pulse alone. The decay length of the noise function is then determined by the width of the front pulse. For decreasing values of $\gw$ both curves approach the value $g=0.64$ (indicated by the red, dashed line), corresponding to the square of the front pulse charge, but then deviate from this for even smaller values of $\gw$.  For the green curve the deviation can be explained as due to charge contribution from the long tail of the front pulse. For the blue curve the oscillations are caused by the charge reflections between the boundaries of the constriction. The damping length of the oscillations  is  determined by the width of the transition region where $g(x)$ changes. {The frequency scale 
is given by the effective velocity $v_c=21$ and the effective length of the constriction $\ell_{\mathrm{eff}}=77$ through $\omega_0=2\pi v_c/\ell_{\mathrm{eff}}$.}  \label{NoiseMiddle}}
\end{figure}

However, in the same way as for the reflected current, this trivial result is caused by multiple reflections between the boundaries of the constriction.  To compensate for the effect of these multiple reflections, which mask the presence of non-integer charges in the limit $\gw\to 0$, it is natural also here to consider the noise for non-vanishing $\gw$. The broadening of the signal, and the reflections between the two boundaries, will affect the low frequency part of the noise, while the form of the front pulse will shape the higher frequency part.
Thus the transmission coefficient $|T(\gw)|$, which tends to 1 for small $\gw$, is expected to approach $\sqrt g$ for large $\gw$. The effect  on the noise is shown in Fig.~\ref{NoiseMiddle}, where the broadening and multiple reflections give rise to oscillations for small $\gw$, while the profile of the noise function for larger $\gw$ is determined by the front pulse. 

In this case the time extension of the front pulse determines the (inverse) width of the noise function, which is therefore much wider than the noise function of the reflected current, and provided the relevant time scales are well separated, the value of the fractional charge can then be extracted for values of $\gw$ where the oscillations are strongly damped. Thus, in this case the value of the fractional charge $\sqrt g$ is (in principle) easier to extract since the signal is much clearer. Note, however that $S_T(\gw)$ depends quadratically on the transmission coefficient $T(\gw)$. The peak value of $S_T(\gw)/(2e\mean{I_{tun}})$ is therefore closer to the quadratic  value $g$ than to the charge value $\sqrt g$ of the front pulse.

\section{Finite temperature effects and background fluctuations} \label{sect V}

In the expressions for the noise used so far, we have implicitly assumed zero temperature, since only tunneling into the Hall bar is assumed. The effect of finite temperature can be taken into account  by assuming that both tunneling into the Hall bar (creation of a charge) and tunneling out of the Hall bar (creation of a hole) can take place, with the relative probability of these two types of events being determined by the Gibbs factor ${\exp(\gb\gD\mu)}$, where $\gD\mu=\mu_2-\mu_1=eV_{tun}$ is the difference in electrochemical potential between the edge of the Hall bar and the tunneling reservoir respectively.
To be more specific, if we assume the temperature to be high enough for the Boltzmann distribution to be valid,  the number
of electrons, $N_+$, and holes,  $N_-$, that are randomly injected into the system are given by 
\be{Npm}
N_+&=&N\frac{\exp(\gb\mu_2)}{\exp(\gb\mu_1)+\exp(\gb\mu_2)}
={N\over 2}\,\frac{\exp(\hf\gb\gD\mu)}{\cosh(\hf\gb\gD\mu)} \nn
N_-&=&N\frac{\exp(\gb\mu_1)}{\exp(\gb\mu_1)+\exp(\gb\mu_2)}  
={N\over 2}\,\frac{\exp(-\hf\gb\gD\mu)}{\cosh(\hf\gb\gD\mu)} \, ,
\ee
corresponding to the mean current
\be{meancur}
\mean{I_{tun}}=e{(N_+-N_-)\over T}=e{N\over T}\tanh(\hf\gb \gD\mu) \, .
\ee
If for simplicity we assume the profile functions for charges and holes to be the same, the expression for $S_{tun}(\gw)$ will however remain unchanged, giving a  temperature dependent relation between the noise and the mean value of the tunneling current,
\be{temp}
S_{tun}(\gw)=2e|f_s(\gw)|^2 \coth(\hf\gb \gD\mu)  \mean{I_{tun}} \, .
\ee
Since  the system is linear, the same factor $\coth(\hf\gb \gD\mu)$ will modify the noise in the reflected current (Eq.\pref{refnoise}) and in the transmitted current {(Eq.\pref{refnoise2})}.

The total noise in the current is then a sum of the background noise, without the tunneling, and the noise in the tunneling current. The background noise is most easily determined in the bosonized theory. Assuming $g(x)$ to be sufficiently smooth, and the velocities sufficiently small, for the pulses to be subject to an adiabatic variation in the value of $g$, the fluctuations are the same as for fixed g. In this case we may use the following momentum expansion of the current, in terms of the bosonic creation and annihilation operators $b_q$ and $b_q^\dag$ \cite{Haldane81}
\be{currentexp}
I(x,t)={e\over L}\sum_q\sqrt{{L g}\over{2\pi |q|}}\, \mbox{sgn}(q)\, \gw_q\,[b_qe^{i(qx-\gw_q t)}+b_q^\dag e^{-i(qx-\gw_q t)}] \, .
\ee
The system is here assumed to be confined to a ring of length $L$ with $g$  as a constant, and the relation between frequency $\gw_q$ and momentum $q$ is $\gw_q=v {|q|}$, with $v$ as the velocity of the edge current. The expression for the noise is then, 
\be{back}
S_0(\gw)&=&\int dt\,  e^{i\gw t} {ge^2\over \pi L}\sum_q {{\gw_q^2}\over {|q|}}\mean{b_q b_q^\dag+ b_q^\dag b_q}\cos(\gw_q t)\nn
&=&{ge^2\over \pi  } {\int_0^{\infty}} d\gw_q \, \gw_q \coth(\hf\gb \gw_q) [\gd(\gw-\gw_q)+\gd(\gw+\gw_q)]\nn
&=&ge^2{\gw\over \pi}  \coth(\hf\gb \gw) \, ,
\ee
where we have taken the continuum limit $L\to \infty$ and assumed a finite temperature Bose-Einstein distribution $\mean{b_q^\dag b_q}=(\exp(\gb\gw_q)-1)^{-1}$. It is interesting to note that the background noise, which is quadratic in the electron charge, is linearly renormalized by the interaction parameter $g$. This means the variation with $g$ is consistent with the picture of an adiabatic change of the electron charge $e\to \sqrt g e$, from the integer to the non-integer value, when the interaction is turned on.

\section{Tunneling within the constriction}
As previously discussed, the fractional charge $\sqrt g \, e$, associated with adiabatically dressed electrons within the constriction, can in principle be detected in a measurement of the current noise in the constriction. Outside the constriction there is an indirect and less clear relation between the charge fractionalization and the noise in the reflected current as well as in the transmitted current. In the transmitted current, outside the constriction, there is in fact no information about the fractional charge in the sharply defined front pulse, since the strength of this is redressed to the value $1$ when $g(x)$ regains this value. Therefore the information lies, as is the case for the reflected current, in the form of the low frequency part of the noise, which depends on how the repeated reflections inside the system shape the signal.

\begin{figure}[h]
\begin{center}
\includegraphics[width=16cm]{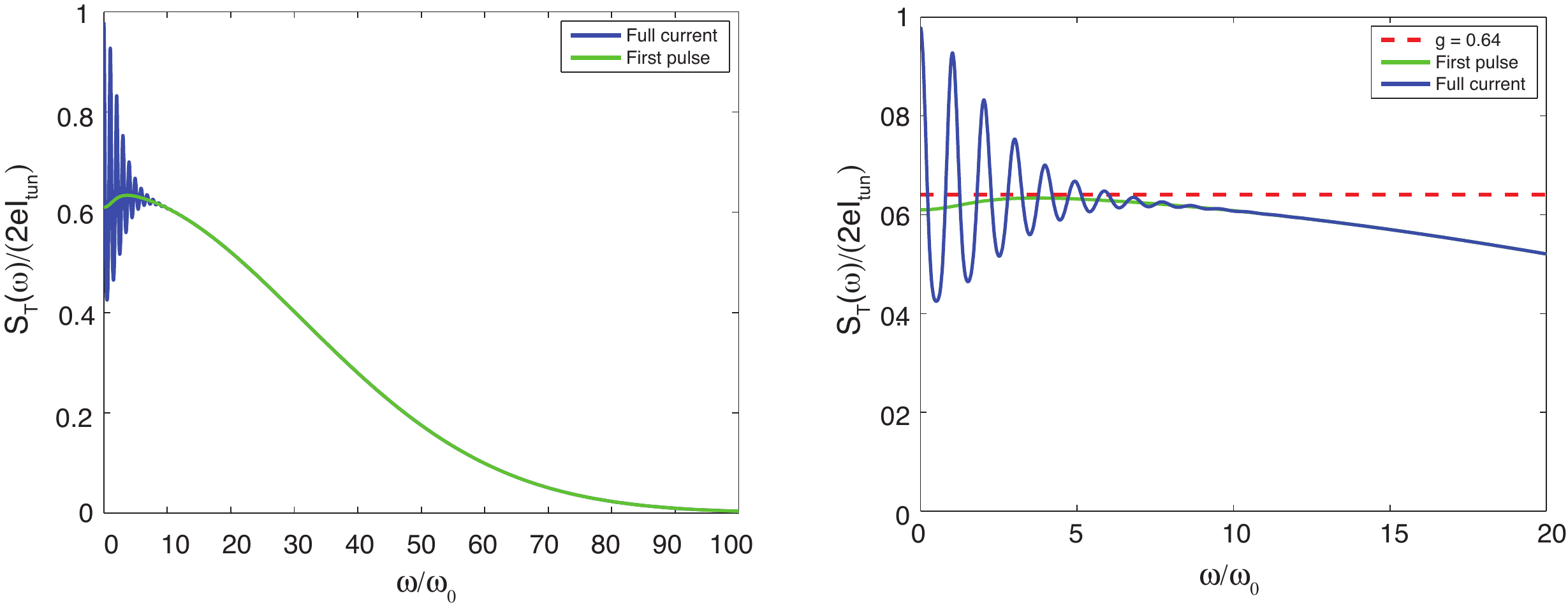}
\end{center}
\caption{\small (Color online) Numerical results for the zero temperature noise outside the constriction, reproduced from simulation of  the charge propagation from one tunneling event within the constriction.  The green curve also here includes only the contribution from the first pulse, and the blue curve is the noise produced by the full current, \ie with secondary pulses included. The noise function is very similar to the noise function  in Fig.\ref{NoiseMiddle} for the case of tunneling taking place on the outside of the constriction. Again, the noise of
the front pulse, for small $\gw$, dips below the value $g=0.64$ (dashed red line), due to contributions from the long tail, and the noise of the full current
shows for small $\gw$ oscillations due to the effects of secondary pulses.  {The frequency scale 
is given by the effective velocity $v_c=21$ and the effective length of the constriction $\ell_{\mathrm{eff}}=117$ through $\omega_0=2\pi v_c/\ell_{\mathrm{eff}}$.} \label{UpperRight}}
\end{figure}

This motivates us to consider a different tunneling scenario. Instead of  coupling to an external reservoir, we assume the distance between the edges of the Hall bar in the constriction to be  sufficiently small to have a non-vanishing probability for charge to tunnel between them.  A potential difference $\gD\mu$ between the two edges will then introduce a tunneling current. As a basic assumption we take that each tunneling event corresponds to moving one electron between the edges. We use the same picture of the tunneling current as before, now with $N_+$ denoting the number of transitions of one electron from the  lower to the upper edge in a given time interval $T$ and $N_-$ as the number of transitions in the opposite direction. With these assumptions the average tunneling current, at non-vanishing temperature, is as before given by Eq.\pref{meancur}. Charge conservation on each edge implies that this is also the average current outside the constriction to the right, while the current to the left is the same in absolute value, but with opposite sign,
\be{right}
\mean {I_{tun}}=\mean {I_{right}}=-\mean {I_{left}}=e{N\over T}\tanh(\hf\gb \gD\mu)
\ee
The expressions for the noise are essentially the same as before, with the noise in the tunneling current given by Eq.\pref{temp},
\be{noisetun}
S_{tun}(\gw)= 2e |f_s(\gw)|^2 \coth(\hf\gb \gD\mu) \mean{I_{tun}}
\ee
with $f_s(\gw)$ again denoting the profile function of the tunneling charge pulse. The noise of the right moving current outside the constriction we write as
\be{noiseright}
S_{right}(\gw)=2e|T_{right}(\gw)|^2 |f_s(\gw)|^2 \coth(\hf\gb \gD\mu) \mean{I_{tun}}
\ee
with $T_{right}(\gw)$ as the Fourier transform of the propagator from the point of tunneling to a point outside the constriction to the right, 
see Fig. \ref{UpperRight}.
The noise in the left moving current is the same. 

To get a qualitative understanding of how these expressions relate to fractional charges within the constriction, we first note that the sudden transition of an electron from the lower to the upper edge will create a right moving pulse within the constriction of strength $g$ and a left moving pulse of strength $-g$. These values are again determined by charge conservation on each edge combined with the fixed ratio between the charges on the two edges for each chiral component. The transmission of this pulse to the outside gives rise to the right moving front pulse with charge determined by multiplication with the inverse dressing factor  $1/ \sqrt g$. Thus the charge is $\sqrt g\, e$, precisely the same charge as for an electron that moves from the non-interacting region into the constriction. The total charge of the right moving pulse is however identical to the electron charge $e$, due to charge conservation on the upper edge. This charge includes the contribution from the tail and the secondary pulses created by reflections within the constriction.

The expected form for the transmission coefficient $T_{right}(\gw)$ is then much the same as for $T(\gw)$, as previously discussed. For small $\gw$ the form is determined by the broadening and multiple reflections within the constriction. Again the limit $\gw\to 0$ gives $T_{right}(0)=1$, consistent with the mean value of the current not depending of $g$, $\mean{I_{right}}=\mean{I_{tun}}$. For larger values of $\gw$  the transmission coefficient approaches the limit $\sqrt g$ determined by the front charge. The result is that the noise shows essentially the same frequency dependence here as in the previous case when measured within the constriction. The main difference is that in this case, with tunneling {\em inside} the constriction, the information about the fractional charge can be extracted by current measurements on the upper edge {\em outside} the constriction, where the measurement can more easily be performed.

\section{Summary and outlook}

In this paper we have shown how an effective Luttinger model, Eq. \pref{action2}, can be used to simulate the low-energy dynamics of charged edge  pulses on a Hall bar with a smoothly varying width. In particular, we have studied the reflection and transmission of such pulses from smooth contractions, described by 
a space dependent Luttinger parameter $g(x)$, and also analyzed the current noise due to tunneling processes. Our  theoretical analysis shows that a charge $e$ pulse transmitted from a non-interacting region ($g=1$) into an interacting region characterized by $g\neq 1$, is composed of a sharp, charge $e \sqrt g$,
 front pulse that suffers no backscattering, and a tail, with a width depending on the velocity and the size of the transition region, due to repeated reflections. 
The reflected pulse, on the contrary, is not expected to show any sharp feature. All these theoretical predictions were confirmed by numerical simulations which clearly show both  the presence of   a well defined $e \sqrt g$ transmitted pulse with a broad tail,  and a broad reflected pulse. We have also analyzed the current noise due to tunneling both from an external reservoir, and within the constriction, using a simple formalism based on the bosonized effective action \pref{action2}. For the former case we concluded that the $e \sqrt g$ charge of the front pulse can be extracted from the noise function $S(\gw)$ provided the latter can be measured at finite $\gw$ {\em inside} the constriction. If the tunneling takes place within the constriction, the fractional charge can be extracted from measurements of $S(\gw)$ {\em outside} the constriction. 

A line of investigation made possible by this work is to use the methods developed in Ref. \onlinecite{LeinaasHorsdal09} to study the properties of quantum  noise due to the propagation of fractionally charged pulses in regions of varying $g$.  The most interesting implication of this work, however,  is the possibility to detect fractional charge $e\sqrt{g}$ by  inducing electron tunneling within a constriction, and measuring the current noise on an external lead. Both the geometry, and the predicted value for the charge differ from earlier proposals. In this paper we have concentrated on the conceptual aspects of the problem, but it is clearly of great interest to  examine if experimental realizations are possible. This may involve more detailed numerical simulations with realistic experimental configurations. In particular, it is important to find out if the effective Luttinger model can describe a realistic experiment.

\vskip 2mm
\noindent
{\bf Acknowledgement} We thank Eddy Ardonne for many discussions and collaboration in an initial stage of this work. We also thank Yuri Galperin for helpful discussions, and support from the Norwegian and Swedish research councils are gratefully acknowledged.


\end{document}